\documentclass[twocolumn,aps,superscriptaddress,prl]{revtex4-1}

\usepackage{graphicx}
\usepackage{dcolumn}
\usepackage{bm}
\usepackage[utf8]{inputenc}
\usepackage{color}
\usepackage{amsmath}
\usepackage{amssymb}
\usepackage{amsbsy}
\usepackage{bbold}
\usepackage{graphicx}
\usepackage{xfrac}
\usepackage{physics}
\usepackage{textcomp}
\usepackage[]{todonotes}
\usepackage[unicode=true,
 bookmarks=false,
 breaklinks=false,pdfborder={0 0 1},backref=false,colorlinks=true]
 {hyperref}
\hypersetup{
 linkcolor=blue,citecolor=blue,urlcolor=blue,linkcolor=blue,citecolor=blue,urlcolor=blue}

\usepackage{soul}
\setstcolor{red}

\begin{document}

\title{Topological States Decorated by Twig Boundary in Plasma Photonic Crystals}

\author{Jianfei Li}
\affiliation{School of Physics, Harbin Institute of Technology, Harbin 150000, People’s Republic of China}

\author{Jingfeng Yao}%
\email{yaojf@hit.edu.cn}
\affiliation{School of Physics, Harbin Institute of Technology, Harbin 150000, People’s Republic of China}
\affiliation{Heilongjiang Provincial Key Laboratory of Plasma Physics and Application Technology, Harbin 150000, People’s Republic of China}
\affiliation{Heilongjiang Provincial Innovation Research Center for Plasma Physics and Application Technology, Harbin 150001, People’s Republic of China}

\author{Ying Wang}%
\affiliation{School of Physics, Harbin Institute of Technology, Harbin 150000, People’s Republic of China}
\affiliation{Heilongjiang Provincial Key Laboratory of Plasma Physics and Application Technology, Harbin 150000, People’s Republic of China}
\affiliation{Heilongjiang Provincial Innovation Research Center for Plasma Physics and Application Technology, Harbin 150001, People’s Republic of China}

\author{Zhongxiang Zhou}%
\affiliation{School of Physics, Harbin Institute of Technology, Harbin 150000, People’s Republic of China}
\affiliation{Heilongjiang Provincial Key Laboratory of Plasma Physics and Application Technology, Harbin 150000, People’s Republic of China}
\affiliation{Heilongjiang Provincial Innovation Research Center for Plasma Physics and Application Technology, Harbin 150001, People’s Republic of China}

\author{Zhihao Lan}%
\email{lanzhihao7@gmail.com}
\affiliation{Department of Electronic and Electrical Engineering, University College London, Torrington Place, London WC1E 7JE, United Kingdom}

\author{Chengxun Yuan}%
\email{yuancx@hit.edu.cn}
\affiliation{School of Physics, Harbin Institute of Technology, Harbin 150000, People’s Republic of China}
\affiliation{Heilongjiang Provincial Key Laboratory of Plasma Physics and Application Technology, Harbin 150000, People’s Republic of China}
\affiliation{Heilongjiang Provincial Innovation Research Center for Plasma Physics and Application Technology, Harbin 150001, People’s Republic of China}

\begin{abstract}

The twig edge states in graphene-like structures are viewed as the fourth states complementary to their zigzag, bearded, and armchair counterparts. In this work, we study a rod-in-plasma system in honeycomb lattice with twig edge truncation under external magnetic fields and lattice scaling and show that twig edge states can exist in different phases of the system, such as quantum Hall phase, quantum spin Hall phase and insulating phase. The twig edge states in the negative permittivity background exhibit robust one-way transmission property immune to backscattering and thus provide a novel avenue for solving the plasma communication blackout problem. Moreover, we demonstrate that corner and edge states can exist within the shrunken structure by modulating the on-site potential of the twig edges. Especially, helical edge states with the unique feature of pseudospin-momentum locking that could be excited by chiral sources are demonstrated at the twig edges. Our results show that the twig edges and interface engineering can bring new opportunities for more flexible manipulation of electromagnetic waves.

\end{abstract}

\maketitle

\textbf{\begin{flushleft}
		1. Introduction
\end{flushleft}}
The realization of topological edge states in photonic systems has attracted much attention due to their unidirectional propagation properties that are immune to disorder, defects, and impurities \cite{wang2009observation,ozawa2019topological,lan2022brief}. With promising potential for the manipulation of electromagnetic waves, photonic topological states have been applied to optical devices such as optical switches, optical integrated circuits and nanolasers to avoid losses caused by large-angle bending and impurities during the preparation process \cite{su2021optical,zhang2020low}. In addition to micro/nanophotonic applications, topological edge states can also be implemented in large-scale scenarios such as communication blackout caused by plasma. The plasma blackout originates from re-entry process of the aircraft into the atmosphere, which provides negative permittivity at a specific frequency and enforces the formation of evanescent waves in plasma \cite{rybak1971progress,li2023observation}. Achieving robust transmission in the negative permittivity background deserves further investigation. Topological states based on photonic crystal platforms in general rely on the structural symmetry and boundary decoration. The widely discussed cases are graphene-like structures, in which Dirac cones exist and topological phase transitions often depend on band inversion at the Dirac point \cite{armitage2018weyl}.

Graphene-like systems feature localized states at the boundaries, which exhibit distinctive transport properties than bulk states. In general, flat band with zero energy occurs at the zigzag edge, which is attributed to the non-binding orbital in the nearest neighbor tight-binding approximation \cite{kohmoto2007zero, ochiai2009photonic}. The bearded edge possesses a flat energy band complementary to the zigzag state, and the Tamm-like mode has been observed near the Van Hove singularity \cite{plotnik2014observation, pantaleon2018effects}. However, there is no edge state for armchair boundary unless defects exist or lattice symmetry breaking occurs \cite{kohmoto2007zero,zheng2019granular,shi2021edge,tang2022valley}. As a magnetic field is applied to a graphene-like system, gapless edge states can be constructed on armchair boundaries, and a series of dispersionless bands emerge corresponding to the Landau levels \cite{lado2022theory}. Recently, a completely new boundary named “twig” has been proposed to realize the localization of light \cite{xia2023photonic}. The twig edge along with the three boundaries mentioned above constitute the four basic boundary types in graphene structures. The substantial developments in topological physics in recent years have facilitated the interest in edge states. It is shown that robust edge states arise from the topological nature of the bulk Hamiltonian, which is summarized as the bulk-edge correspondence \cite{graf2013bulk,xiao2014surface}. 

When the time-reversal (TR) symmetry is preserved, a well known Wu-Hu model is proposed to realize photonic helical edge states by constructing pseudospin $ 1/2 $ states \cite{wu2015scheme}. This well simulates the quantum spin Hall (QSH) effect in electronic systems utilizing all-dielectric photonic crystals. The nontrivial nature of the Wu-Hu model has been demonstrated in \cite{palmer2021berry} by Palmer and Giannini, who showed that considering the pseudospin supported by the $ C_2T $ symmetry of the Wu-Hu model, one can obtain $ \pm 2\pi $ phase winding in the Wilson loops for the two pseudospin channels. They further clarified that the so-called fragile topology in \cite{de2019engineering} can be adiabatically connected to the Wu-Hu model without changing the ${C_{6v}}$ symmetry and closing the band gap, indicating that the two models are the same in topology and thus share the same helical topological edge states. Moreover, Kariyado and Hu in \cite{kariyado2017topological} demonstrated that graphene-zigzag bulk unit cell hosts a nonzero winding number leading to dispersionless bands in the band gap and the partially-bearded interfaces have dispersive edge states which are protected by a pair of mirror winding numbers. Based on Wu-Hu model, scaling or rotation of circular rings and elliptical cylinders is proposed to realize the QSH effect in photonic systems \cite{xu2016accidental,jiang2019manipulation,zhou2021topological}. Between two photonic crystals with different topological properties, unidirectional edge states could be excited at the interface by chiral sources. The tunable directionality and robust properties of the Wu-Hu model extended to fabricate optical circuits, large-area waveguides, and topological lasing \cite{kim2020recent,lan2023large,song2023observation}.

In this work, we develop a rod-in-plasma system to investigate twig edge behaviors in honeycomb lattice by simultaneously breaking TR symmetry and pseudo-TR symmetry, which is different from the rod-in-air systems, where edge states exist at the interface between two structures with different topological properties, or below the light cone without cladding structure \cite{poo2011experimental,chen2020local}. Under a magnetic field, the plasma photonic crystal (PPC) undergoes a phase transition from conventional insulator to Chern insulator and then to QSH insulator by changing the positions of the rods and the strength of the magnetic field. The realization of robust transmission in negative permittivity background paves the way to solve the plasma blackout problem. By modulating the on-site potential of the twig boundary, corner states can appear in the band gap. These tunable properties also have promising applications such as spin splitter devices, resonators and highly sensitive sensors \cite{wang20193d,chen2017multiple,liu2022tunable}.

\begin{figure}[t]
\begin{center}
\includegraphics[width=0.49\textwidth]{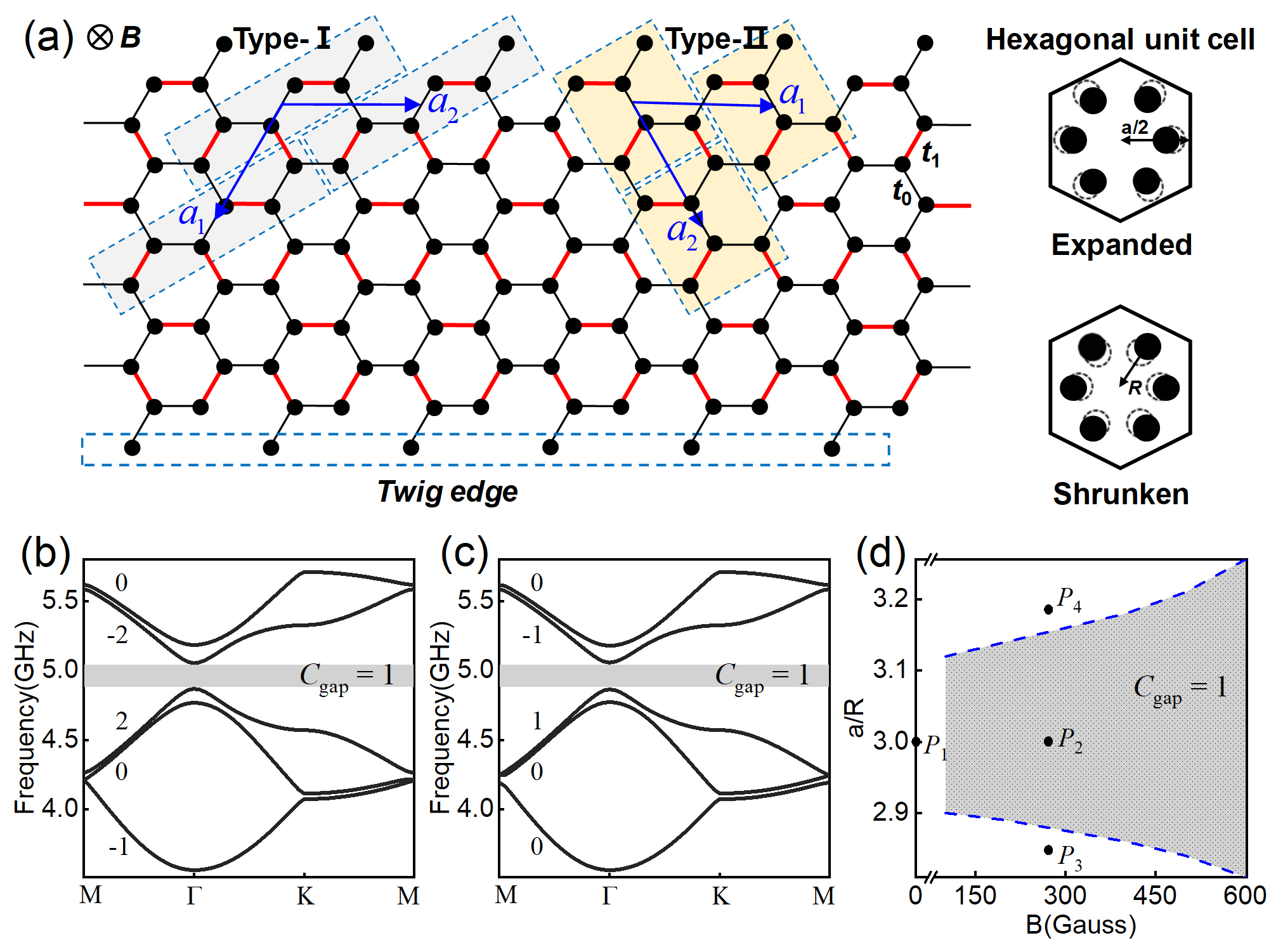}
\end{center}
\caption{\label{figs:fig1} The model of plasma photonic crystal with twig edges and the calculated band diagrams under expanded and shrunken cases. (a) PPC constructed by YIG rods immersed in plasma background with a magnetic field. The band diagram when the rods are (b) expanded by $R = a/2.95 $ and (c) shrunken by $R = 3.05a$. Nontrivial band gaps with ${C_{\text{gap}}} = 1$ are obtained for both cases. (d) The phase diagram of PPC at different rod positions and magnetic fields.}
\end{figure}

\textbf{\begin{flushleft}
		2. Phase transition of graphene-like structure in plasma
\end{flushleft}}

The plasma photonic crystals is depicted in Fig. 1(a), where yttrium iron garnet (YIG) rods arranged in a graphene-like lattice are imbedded in plasma background and each unit cell contains six YIG rods (lattice constant $a = 40$ mm, rod radius $r = 0.1a$, the distance between rod and center of unit cell is $ R $, electron density ${n_e} = 1 \times {10^{12}}cm{^{ - 3}}$). The relative permittivity of plasma is ${\varepsilon _r} = 1 - {{\omega _{pe}^2} \mathord{\left/{\vphantom {{\omega _{pe}^2} {{\omega ^2}}}} \right.\kern-\nulldelimiterspace} {{\omega ^2}}}$ (plasma frequency is expressed as $ {\omega _{pe}} = {{n{e^2}}\mathord{\left/{\vphantom {{n{e^2}} {{\varepsilon _0}}}}\right.\kern-\nulldelimiterspace}{{\varepsilon _0}}}{m_e} $, $ n $ represents the electron density, $ e $ the unit charge, ${\varepsilon _0}$ the permittivity in vacuum, and ${m_e}$ the electron mass), where the collision frequency is neglected for simplicity. The plasma permittivity is negative when the incident wave frequency is lower than the plasma frequency. Permeability of the YIG rods could be expressed as a tensor when the external magnetic field is perpendicular to the structure, which leads to the breaking of TR symmetry \cite{wang2008reflection}. When damping coefficient is neglected, the permeability of YIG is expressed as,
\begin{equation}
\hat \mu {\rm{ = }}\left[ {\begin{array}{*{20}{c}}
	\mu &{ - i{\mu _1}}&0\\
	{i{\mu _1}}&\mu &0\\
	0&0&1
	\end{array}} \right]
\end{equation}
with $\mu $ and ${\mu _1}$ given by
\begin{equation}
\mu  = 1 + \frac{{{\omega _m}{\omega _0}}}{{{\omega _0}^2 - {\omega ^2}}},{\rm{ }}{\mu _1} = \frac{{{\omega _m}\omega }}{{{\omega _0}^2 - {\omega ^2}}}
\nonumber
\end{equation}
The resonant frequency $ \omega _0 $ depends on the strength of the external magnetic field, which is expressed as ${\omega _0} = 2\pi \gamma {H_0}$ ($\gamma  = 2.8$ MHz/Oe). ${{\omega _m}}$ is the coupling strength of the YIG material to the magnetic field, which is taken as $3.08 \times {10^{10}}$ rad/s \cite{liu2011molding}. The magnetized plasma has similar properties when the magnetic field is parallel to the structure, which can serve as a topological material \cite{gao2016photonic,parker2020topological,fu2021topological}, and the magnetized plasma is usually realized by semiconductor materials \cite{wu2018topologically,wu2023strong}.The whole structure is terminated with twig-shaped edges that are viewed as the fourth type of edges of the honeycomb lattice \cite{xia2023photonic}.

We consider the case when the magnetic field is perpendicular to the structure and transverse magnetic (TM, ${E_z},{H_x},{H_y}$) modes of the electromagnetic wave will be considered in the following. The bands we focus on are below the plasma cutoff frequency, i.e., the YIG rods are surrounded by negative permittivity plasma. All the simulation results in this article are performed by the finite element method implemented within the COMSOL Multiphysics software, and the details can be found in S1 of Supporting Information. There are three types of unit cells that can be chosen, i.e. Hexagonal unit cell, Type-\uppercase\expandafter{\romannumeral1} and Type-\uppercase\expandafter{\romannumeral2} zigzag bulk unit cell. When the time-reversal symmetry is broken, they share the same band structures and Chen numbers as shown in S2 of Supporting Information. The Hexagonal unit cell is first selected to clearly show the shrinking and expanding features. We fix the magnetic field strength at 300 Gauss, and slightly expand ($R = {a \mathord{\left/{\vphantom {a {2.95}}} \right.\kern-\nulldelimiterspace} {2.95}}$) the six rods within a unit cell. The band diagram is shown in Fig. 1(b), from which one can see an omnidirectional band gap between the third and fourth bands. As the TR symmetry of the system is broken, the bands could have different topological properties, which can be characterized by the Chen number. The Chern number of each non-degenerate band is given by \cite{ochiai2009photonic,wang2020universal,zhao2020first},\par

\begin{equation}
{C_n} = \frac{1}{{2\pi }}\iint_{BZ} {{\nabla _k}}  \times {{\rm{A}}_{nk}}{d^2}k
\end{equation}
where ${{\rm{A}}_{nk}}$ is the Berry connection and can be expressed as
\begin{equation}
{A_{nk}} =  - i\left\langle {{u_{nk}}} \right|{\nabla _k}\left| {{u_{nk}}} \right\rangle
\end{equation}
with ${u_{nk}}$ the spatially periodic part of the Bloch function of the $ n $th band. By discretizing the Brillouin zone, the Chen numbers of the expanded lattice are calculated as -1, 0, 2, -2, 0 from the lower band to the upper band (see the marked Chern number for each band in Fig. 1(b)). The non-trivial nature of the bandgap is then evident by ${C_{\text{gap}}} = 1$. Next, we shrink the six rods within a unit cell by $R = 3.05a$, and the results are shown in Fig. 1(c). The shrunken case shares roughly the same band gap with Fig. 1(b), and the Chern number of the band gap is also ${C_{\text{gap}}} = 1$. Moreover, the topological properties of the bands can also be determined by their parities at the $\Gamma $ point. The order of the eigenstates [${p_ \pm },{d_ \pm }$] on either side of the band gap determines the QSH phase, the quantum Hall (QH) phase or the insulating phase of the system \cite{chen2017multiple}. The phase diagram of the plasma photonic crystal under different magnetic fields and shrinking/expanding scales is obtained as shown in Fig. 1(d). The shaded area corresponds to the QH phase with ${C_{\text{gap}}} = 1$. Below the shaded region of the phase diagram, the system is in the QSH phase. As the six rods within the unit cell continue to shrink, the system transforms into an insulating phase with trivial band gap (${C_{gap}} = 0$). We will choose four points in Fig. 1(d) (i.e., $P_1, P_2, P_3$ and $ P_4$) to analyze the properties of the twig edge states.

\begin{figure}[t]
	\includegraphics[width=0.48\textwidth]{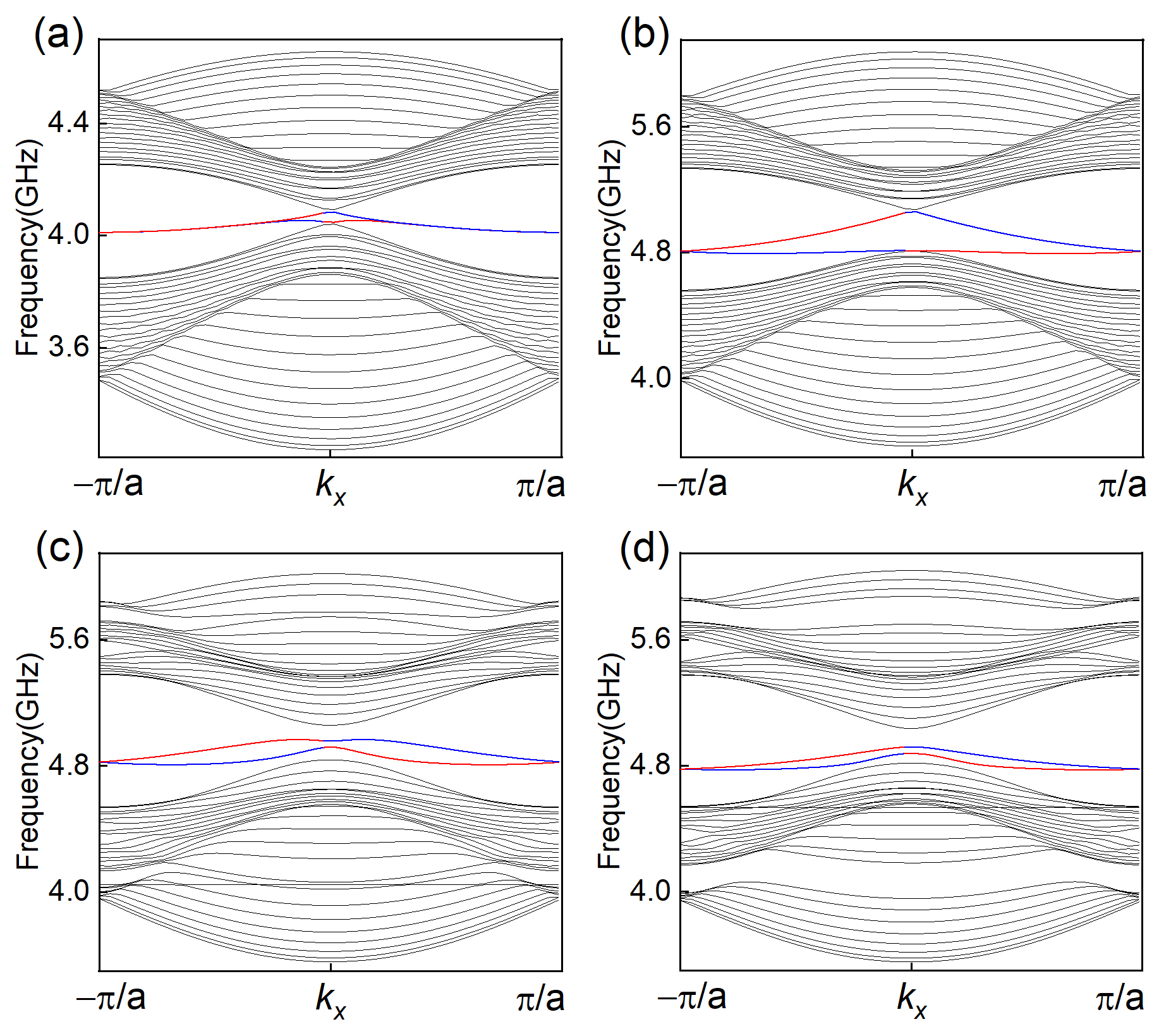}
	\caption{\label{figs:fig2} The projected band diagram along the $x$ direction for (a) ${P_1}$, (b) ${P_2}$, (c) ${P_3}$, and (d) ${P_4}$ as marked in Fig. 1(d). The results show that gapless topological edge states exist in the nontrivial band gap, and the gapped edge states can be found both in expanded and shrunken cases. }
\end{figure}

To demonstrate the influence of twig edge on the non-bulk states, we select a supercell from Fig. 1(a) and calculate the band diagram projected in the $ x $ direction. In the absence of a magnetic field and lattice scaling, a full flat-band mode was found with nontrivial topological winding number \cite{xia2023photonic}. Similar result can be observed for PPC as shown in Fig. 2(a), which corresponds to ${P_1}$ in Fig. 1(d). A nearly flat band appears across the Brillouin zone, and the lower twig edge state (red line) degenerates with the upper one (blue line). When the TR symmetry is broken as described at ${P_2}$, the degenerate twig edge states undergo splitting. A gapless edge state appears on each twig interface, corroborating the bulk-edge correspondence. For the case of ${P_3}$, there are gapped edge states in the pseudo gap, and a discrete band spectrum appears near ${k_x} = 0$ [see Fig. 2(c)]. Different from the topological gapless edge states in the QSH system \cite{lado2022theory,chen2017multiple}, the twig edge states are pushed into the band gap. It depends strongly on the on-site boundary potential of the twig edge \cite{yao2009edge}, and the sizes of the rods at the twig edge enable the edge states to couple to the upper or lower bulk bands. As for the shrunken case described by ${P_4}$, the edge states also appear in the band gap as shown in Fig. 2(d).

To explain the nature of the twig edge states in the shrunken case, we start from the tight-binding method to describe the graphene-like structure. The Hamiltonian is described by $H{\rm{ = }}\sum\limits_{\left\langle {i,j} \right\rangle } {\left( {{t_{ij}}c_i^\dag {c_j} + h.c.} \right)} $, where $c_i^\dag $ and ${c_j}$ are the creation and annihilation operators at sites $ i $ and $ j $. ${t_{ij}}$ is the hopping amplitude between lattice sites, and we define the intracell hopping strength as ${t_0}$ and the intercell hopping strength as ${t_1}$ [see FIG. 1(a)]. In this case, the Type-\uppercase\expandafter{\romannumeral1} zigzag bulk unit cell is considered since the bulk unit cell and the shape of edge are tied up \cite{kariyado2017topological}, i.e., the unit cell is not broken at the twig edge. As for the Type-\uppercase\expandafter{\romannumeral1} zigzag bulk unit cell, the Hamiltonian is written as
\begin{equation}
{H_k} = \left( {\begin{array}{*{20}{c}}
	0&{{Q_k}}\\
	{Q_k^\dag }&0
	\end{array}} \right)
\end{equation}
where ${Q_k}$ is $3 \times 3$ matrix of,
\begin{equation}
{Q_k} = \left( {\begin{array}{*{20}{c}}
	{{t_0}{e^{ - i\bm{k} \cdot \bm{a_2}}}}&{{t_1}{e^{ - i\bm{k} \cdot \left( \bm{{a_2} - {a_1}} \right)}}}&{{t_0}}\\
	{{t_0}{e^{ - i\bm{k} \cdot \left( \bm{{a_2} - {a_1}} \right)}}}&{{t_0}}&{{t_1}}\\
	{{t_1}}&{{t_0}}&{{t_1}{e^{ i\bm{k} \cdot \bm{a_1}}}}
	\end{array}} \right)
\end{equation}
Then, the winding number of the system can be obtained by the following equation\cite{ryu2010topological,kariyado2017topological},
\begin{equation}
n\left( {{k_\parallel }} \right) =  - \frac{1}{{2\pi }}\int_0^{2\pi } {\frac{d}{{d{k_ \bot }}}} \arg \left( {\det Q_k} \right)d{k_ \bot }
\end{equation}
Due to the symmetry of the zigzag bulk unit cell, we focus on the case $ {k_\parallel } = 0 $ and obtain that the winding number of the shrunken case is 1 (for details, see S3 of the Supporting Information). We would like to note that when studying edge states in the context of band topology based on crystalline symmetry, caution should be taken because whether a band gap is topologically trivial or nontrivial relies heavily on the choice of the bulk unit cell. As Kariyado and Hu have pointed out in \cite{kariyado2017topological}, choice of the unit cell should be made such that it is not broken at the considered edge. Indeed, the hexagonal unit cell, which in the literature is considered as trivial for the shrunken lattice, is not appropriate for the study of the twig edge due to its broken nature at the edge. Instead, the two kinds of zigzag unit cells defined in Fig.\ref{figs:fig1}(a) and intact at the edge should be used, and our results show that the zigzag unit cell  has a nontrivial winding number for both the shrunken and expanded lattices.

\begin{figure}[b]
	\includegraphics[width=0.49\textwidth]{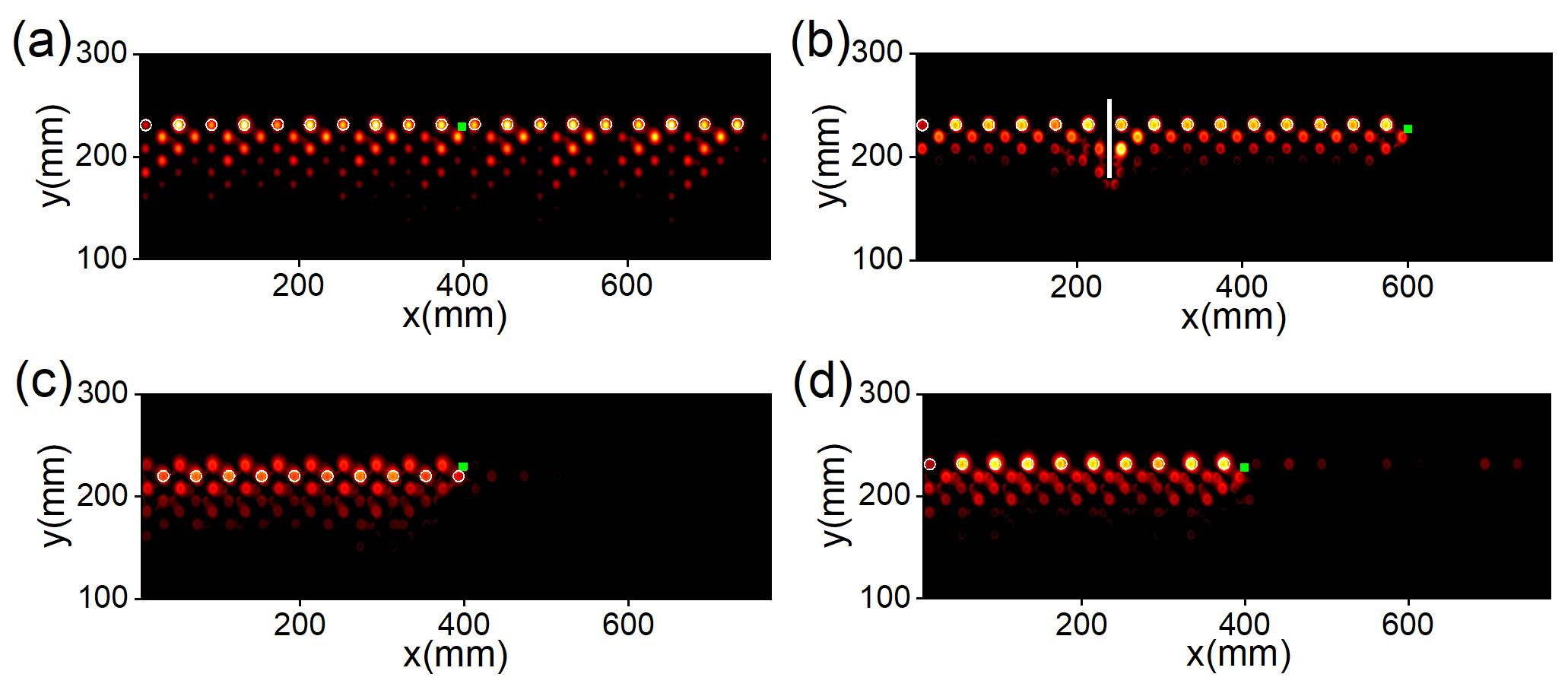}
	\caption{\label{figs:fig3} Full-wave simulations of twig edge states excited by a line source for different cases: (a) magnetic field and lattice scaling are absent, which demonstrates the confinement of energy at the twig interface; (b) TR symmetry is broken, revealing the robustness of unidirectional transmission; (c) lattice expansion and magnetic fields are present, where mixed modes are demonstrated; (d) lattice contraction and magnetic fields are present, which exhibits a unidirectional propagation of twig edge states in a band gap with zero Chern number but nonzero winding number.}
\end{figure}\par

We further perform full-wave simulations to verify the four twig edge modes. The YIG rods corresponding to the four situations are imbedded in plasma background, and a line source is placed at the twig interface. When the lattice scaling and magnetic field are absent for the case of $P_1$, the energy is localized at the twig interface and penetrates slightly into the interior of the PPC as shown in Fig. 3 (a). As the TR symmetry is broken for $P_2$, unidirectional transmission can be observed in Fig. 3 (b) at 4.87 GHz. Importantly, the twig edge state in this case is robust to impurities and free from backward scattering. By this way, we realize the robust transmission of electromagnetic waves in the negative permittivity background, which provides a new approach to communication in the plasma blackout. As for the expanded case at $P_3$, the twig edge state excited at 4.93 GHz is shown in Fig. 3 (c). One can see a clear mixed mode here in that the energy is predominantly localized to the inner rods of the twig interface. Interestingly, the twig edge mode at $P_4$ can also be unidirectionally excited in the insulating phase at 4.84 GHz [Fig. 3(d)]. When there is no external magnetic field, the twig edge state with non-zero winding number is approximately a flat band in the gap. However, surface plasmon resonance induced by magnetic fields produces the sharply asymmetric reflection effect \cite{liu2011molding}. This effect results in dispersive twig bands, allowing electromagnetic waves to travel in one direction. If the magnetic field is increased to 2000 Oe, the resonance frequency of the surface plasmon deviates greatly from the twig edge state. Then, the twig state returns to the dispersionless situation (see Fig. S6 of the Supporting Information for the band diagram and transmission spectrum in this case).

\textbf{\begin{flushleft}
		3. Corner and helical edge states in shrunken case with nontrivial winding number
\end{flushleft}}

The on-site potential of the boundary can dramatically affect the dispersion characteristics of the edge states \cite{xi2021observation}, which could bring novel effects for the twig edges. To demonstrate this, we consider the shrunken case as illustrated in Fig. 2(d). As the size of the boundary atoms decreases, the edge states are gradually pushed into the upper bulk bands. On the other hand, the edge states will merge into the lower bulk bands when the size of the boundary atoms increases. The reason for this could be attributed to the fact that sufficiently small or large on-site potentials of the twig edges will lead to the decoupling of the twig edges from neighboring rods and as such, the system effectively terminates with armchair edges, where a complete band gap develops. When the size of the boundary atoms is fixed to $0.092a$, two band gaps named gap 1 and gap 2 appear near the edge states as shown in Fig. 4(a), in which corner states are expected. To show this, we construct a triangular-shaped supercell terminated with twig edges and containing 14-unit cells per edge, which is immersed into the plasma. The calculated eigenmodes of the supercell are shown in Fig. 4(b), from which we can find that corner states marked by red triangles exist in the two band gaps. The black and blue dots represent bulk and edge states respectively. For the corner states in gap 1, their electric fields (${E_z}$) are localized at three different corners, respectively [see Fig. 4(d)] whereas the fields of corner states in gap 2 are distributed to three corners simultaneously (results not shown here). When small defects are introduced, e.g., one of the rods around the corner is removed, the spectrum differs remarkably as shown in Fig. 4(c), where the original corner states in gap 1 disappear, while the corner states in gap 2 are merged into the edge states. The field distributions of the merged edge states are shown in Fig. 4(e), from which one can see that the energy is mainly localized near the corners and then spreads towards the center of the edge. The topological property, which is characterized by the non-zero winding number, of the zigzag bulk unit cell depends on the symmetry of the structure. Defects could break the symmetry of the zigzag bulk unit cell, which causes the corner states to disappear.

\begin{figure}[t]
	\includegraphics[width=0.5\textwidth]{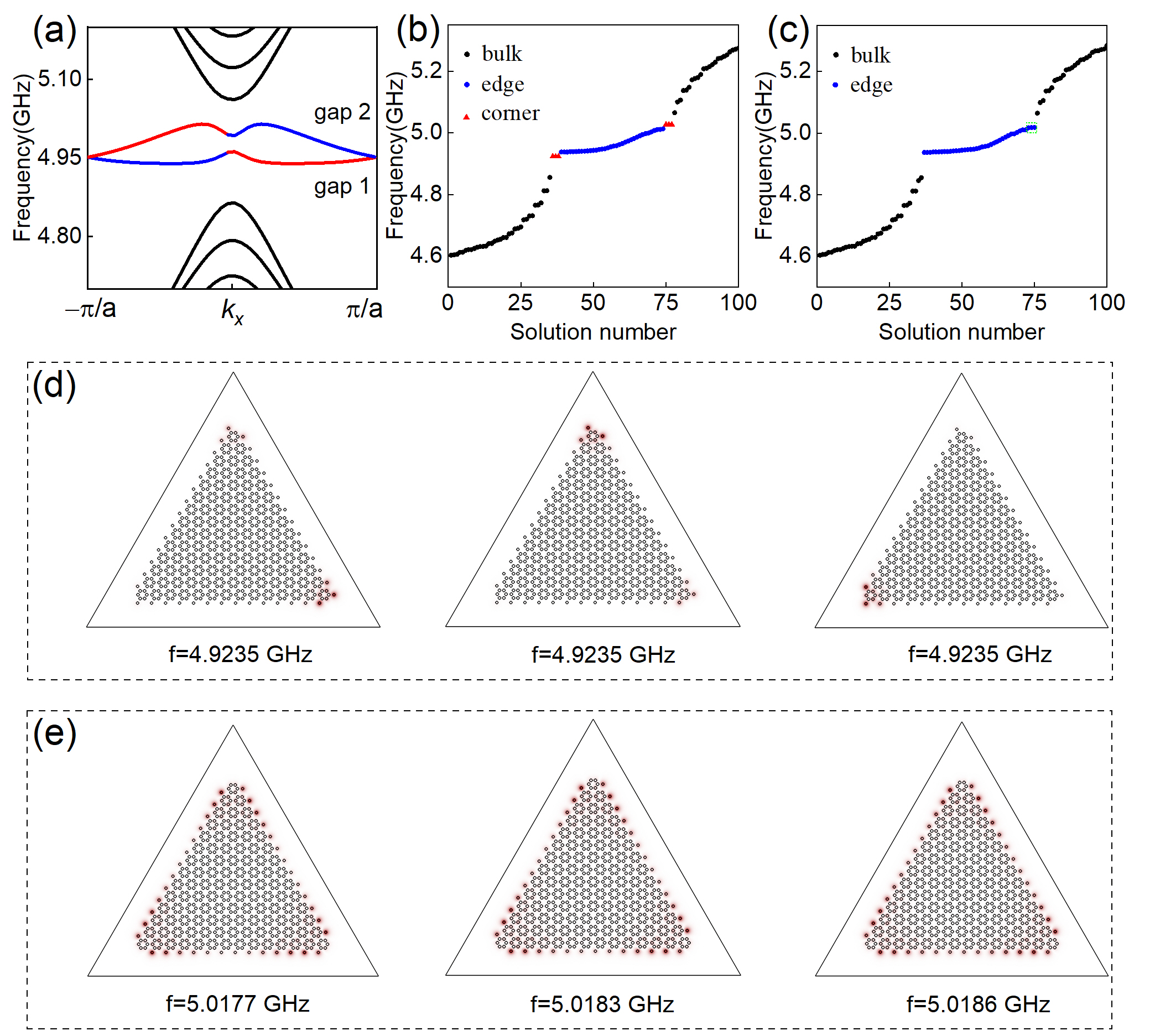}
	\caption{\label{figs:fig4} Corner states in shrunken lattice and modulated by twig interface. (a) The band diagram projected in the x direction for shrunken lattice with boundary atoms of size $0.092a$. By modulating the on-site boundary potential, the gapped edge states will be pushed into either the upper or lower bulk bands. (b) Calculated eigenmodes of a triangular-shaped supercell, where the black and blue dots represent the bulk and edge states respectively, whereas the red triangles denote the corner states. (c) The calculated eigenmodes when defects are present. (d) The field distributions (${E_z}$) of corner states in gap 1. (d) The field distributions of edge states marked within the green box in Fig. 4(c).}
\end{figure}

\begin{figure}[t]
	\includegraphics[width=0.49\textwidth]{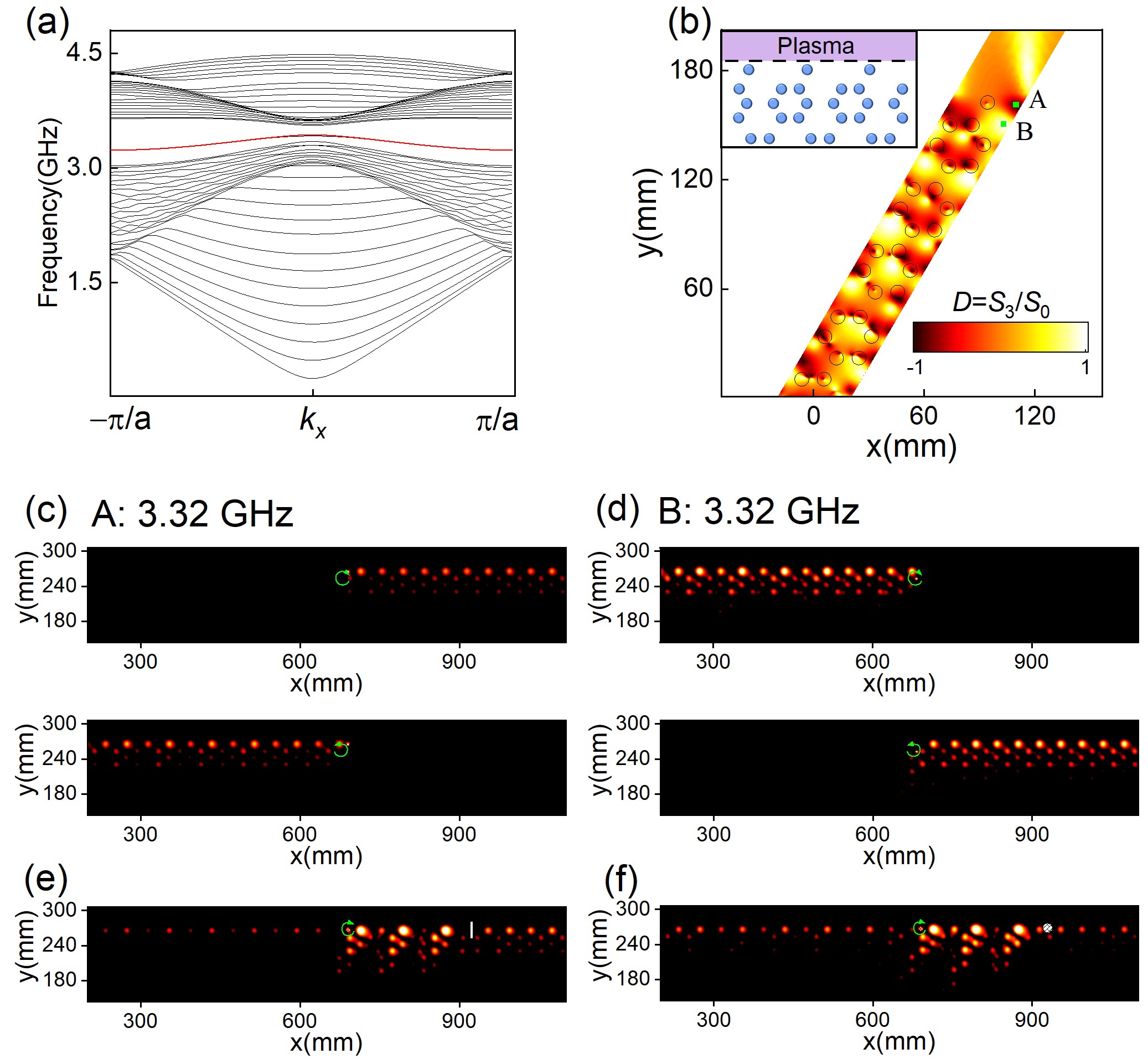}
	\caption{\label{figs:fig5} Helical edge states at the twig edge of a shrunken lattice. (a) The band diagram projected in the x direction, which shows a dispersive band of edge states within the band gap. (b) The schematic of the twig edge structure cladded by plasma, and the chirality map defined by the Stokes parameters ${S_0}$ and ${S_3}$. Full-wave simulations of wave excited by the chiral source at the location marked by (c) A, and (d) B. (e) A rectangular impurity is placed along the propagation path. (f) A small defect is introduced at the boundary.}
\end{figure}

The invention of twig edge not only discovers a new localization mode \cite{xia2023photonic}, but also serves as a useful paradigm for interface engineering to explore the unconventional properties in topological physics. Here we consider a shrunken lattice with $R=a/3.15$,  where the rods are placed into air without an external magnetic field, while the twig edge is enveloped in plasma [see the inset of Fig. 5 (b)]. The calculated projected band structure is shown in Fig. 5(a). A dispersive band of edge states appears in the band gap due to the twig interface, which is not observed in structures with armchair edges. The local chirality of twig edge state can be characterized as $D = {{{S_3}} \mathord{\left/{\vphantom {{{S_3}} {{S_0}}}} \right.\kern-\nulldelimiterspace} {{S_0}}}$ in terms of the Stokes parameter of the magnetic field, where ${S_0} = {\left| {{H_x}} \right|^2} + {\left| {{H_y}} \right|^2}$ and ${S_3} =  - 2{\mathop{\rm Im}\nolimits} \left( {{H_x}H_y^*} \right)$ \cite{lan2021second}. The chirality map of the edge state at ${k_x} = 0.48{\pi  \mathord{\left/{\vphantom {\pi  a}} \right.\kern-\nulldelimiterspace} a}$ is shown in Fig. 5(b), from which we can find distinct negative and positive chirality near the twig edge. As the chiral source is placed at the dark-red (point A) or bright yellow (point B) region, the directionality of electromagnetic waves may exhibit dissimilar property. Full-wave simulations for exploring the transmission properties are presented in Figs. 5(c) and (d). The chiral source constructed by four dipoles with clockwise or counterclockwise phases is placed at the negative chirality region marked by A. When the frequency is fixed at 3.32 GHz with clockwise phases for the chiral source, the mode with unidirectional propagation to the right is excited. As for the counterclockwise counterpart, the electromagnetic wave propagates unidirectionally to the left along the interface as shown in Fig. 5(c). The behavior will be reversed when the chiral source is placed at B, i.e., wave unidirectionally propagating to the left will be excited by the chiral source with clockwise phases and vice versa [see Fig. 5(d)]. We then place a small rectangular impurity along the propagation path as shown in Fig. 5 (e). The electromagnetic wave can pass through this impurity, though the intensity is attenuated behind the impurity. The twig edge state can also pass through a defect though with a certain degree of reflection as shown in Fig. 5 (f). The main reason for this phenomenon is that the twig edge state originates from the zigzag bulk unit cell, and defects will cause the completeness of the unit cell to be broken. Though the twig edge state in the shrunken lattice is not as robust as that of the expanded case (see Fig. S7 of the Supporting Information), this boundary decoration approach can provide another degree of freedom for modulating electromagnetic waves. Moreover, in practice, the robustness of the edge states against moderate disorder is beneficial for their experimental realizations, where perfect edges in real applications can not be guaranteed due to the inevitable impurities and defects created during device fabrication.
\textbf{\begin{flushleft}
		4. Conclusion
\end{flushleft}}

We have developed the rod-in-plasma system, and studied the twig edge states under external magnetic field and lattice scaling. When the TR symmetry is broken, topological edge states exist at the twig interface, which can be maintained in limited shrunken and expanded lattices. The demonstration of robust transmission of electromagnetic waves in a negative permittivity background paves the way for solving the communication blackout problem in plasma. We have demonstrated that twig edge states can exist in the extremely shrunken case, which stems from the non-zero winding number of the zigzag bulk unit cell. And the twig state is influenced by surface plasmon resonance, resulting in the unidirectional transmission. The corner states can also be constructed in shrunken structure when terminating the system with twig edges. In system preserving the TR symmetry, helical edge state excitable by chiral sources are formed due to the twig interface. The transmission direction can be changed when the source is placed at a location with different chirality.\\

\textbf{\begin{flushleft}
		Supporting Information
\end{flushleft}}
Supporting Information is available from the Wiley Online Library or from the author.

\textbf{\begin{flushleft}
		Acknowledgments
\end{flushleft}}
This work was supported by the Natural Science Foundation of China (Nos. 12175050 and 12205067) and the Fundamental Research Funds for the Central Universities (Grant No. HIT. OCEF. 2022036 and HIT. DIJJ. 2023178).

\textbf{\begin{flushleft}
		Conflict of Interst
\end{flushleft}}
The authors declare no conflict of interest.

\textbf{\begin{flushleft}
		Data Availability Statement
\end{flushleft}}
The data that support the findings of this study are available from the corresponding author upon reasonable request.

\section*{References}

\begin{thebibliography}{10}
	\providecommand{\url}[1]{\texttt{#1}}
	\providecommand{\urlprefix}{URL }
	
	\bibitem{wang2009observation}
	Z.~Wang, Y.~Chong, J.~D. Joannopoulos, M.~Solja{\v{c}}i{\'c},
	\newblock \emph{Nature} \textbf{2009}, \emph{461}, 7265 772.
	
	\bibitem{ozawa2019topological}
	T.~Ozawa, H.~M. Price, A.~Amo, N.~Goldman, M.~Hafezi, L.~Lu, M.~C. Rechtsman,
	D.~Schuster, J.~Simon, O.~Zilberberg, et~al.,
	\newblock \emph{Rev. Mod. Phys.} \textbf{2019}, \emph{91}, 1 015006.
	
	\bibitem{lan2022brief}
	Z.~Lan, M.~L. Chen, F.~Gao, S.~Zhang, E.~Wei,
	\newblock \emph{Rev. Phys.} \textbf{2022}, 100076.
	
	\bibitem{su2021optical}
	R.~Su, S.~Ghosh, T.~C. Liew, Q.~Xiong,
	\newblock \emph{Sci. Adv.} \textbf{2021}, \emph{7}, 21 eabf8049.
	
	\bibitem{zhang2020low}
	W.~Zhang, X.~Xie, H.~Hao, J.~Dang, S.~Xiao, S.~Shi, H.~Ni, Z.~Niu, C.~Wang,
	K.~Jin, et~al.,
	\newblock \emph{Light Sci. Appl.} \textbf{2020}, \emph{9}, 1 109.
	
	\bibitem{rybak1971progress}
	J.~P. Rybak, R.~Churchill,
	\newblock \emph{IEEE Trans. Aerosp. Electron. Syst.} \textbf{1971}, , 5 879.
	
	\bibitem{li2023observation}
	J.~Li, J.~Yao, Y.~Wang, Z.~Zhou, A.~A. Kudryavtsev, Z.~Lan, C.~Yuan,
	\newblock \emph{APL Photonics} \textbf{2023}, \emph{8}, 6.
	
	\bibitem{armitage2018weyl}
	N.~Armitage, E.~Mele, A.~Vishwanath,
	\newblock \emph{Rev. Mod. Phys.} \textbf{2018}, \emph{90}, 1 015001.
	
	\bibitem{kohmoto2007zero}
	M.~Kohmoto, Y.~Hasegawa,
	\newblock \emph{Phys. Rev. B} \textbf{2007}, \emph{76}, 20 205402.
	
	\bibitem{ochiai2009photonic}
	T.~Ochiai, M.~Onoda,
	\newblock \emph{Phys. Rev. B} \textbf{2009}, \emph{80}, 15 155103.
	
	\bibitem{plotnik2014observation}
	Y.~Plotnik, M.~C. Rechtsman, D.~Song, M.~Heinrich, J.~M. Zeuner, S.~Nolte,
	Y.~Lumer, N.~Malkova, J.~Xu, A.~Szameit, et~al.,
	\newblock \emph{Nat. Mater.} \textbf{2014}, \emph{13}, 1 57.
	
	\bibitem{pantaleon2018effects}
	P.~A. Pantale{\'o}n, Y.~Xian,
	\newblock \emph{J. Phys. Soc. Japan} \textbf{2018}, \emph{87}, 6 064005.
	
	\bibitem{zheng2019granular}
	L.-Y. Zheng, F.~Allein, V.~Tournat, V.~Gusev, G.~Theocharis,
	\newblock \emph{Phys. Rev. B} \textbf{2019}, \emph{99}, 18 184113.
	
	\bibitem{shi2021edge}
	Z.~Shi, M.~Zuo, H.~Li,
	\newblock \emph{Results Phys.} \textbf{2021}, \emph{24} 104191.
	
	\bibitem{tang2022valley}
	Q.~Tang, M.~R. Beli{\'c}, Y.~Q. Zhang, Y.~P. Zhang, Y.~D. Li,
	\newblock \emph{Nonlinear Dyn.} \textbf{2022}, \emph{108}, 2 1573.
	
	\bibitem{lado2022theory}
	J.~Lado, J.~Fern{\'a}ndez-Rossier,
	\newblock \emph{arXiv preprint arXiv:2210.07568} \textbf{2022}.
	
	\bibitem{xia2023photonic}
	S.~Xia, Y.~Liang, L.~Tang, D.~Song, J.~Xu, Z.~Chen,
	\newblock \emph{Phys. Rev. Lett.} \textbf{2023}, \emph{131} 013804.
	
	\bibitem{graf2013bulk}
	G.~M. Graf, M.~Porta,
	\newblock \emph{Commun. Math. Phys.} \textbf{2013}, \emph{324} 851.
	
	\bibitem{xiao2014surface}
	M.~Xiao, Z.~Zhang, C.~T. Chan,
	\newblock \emph{Phys. Rev. X} \textbf{2014}, \emph{4}, 2 021017.
	
	\bibitem{wu2015scheme}
	L.-H. Wu, X.~Hu,
	\newblock \emph{Phys. Rev. Lett.} \textbf{2015}, \emph{114}, 22 223901.
	
	\bibitem{palmer2021berry}
	S.~J. Palmer, V.~Giannini,
	\newblock \emph{Phys. Rev. Res.} \textbf{2021}, \emph{3}, 2 L022013.
	
	\bibitem{de2019engineering}
	M.~B. De~Paz, M.~G. Vergniory, D.~Bercioux, A.~Garc{\'\i}a-Etxarri, B.~Bradlyn,
	\newblock \emph{Phys. Rev. Res.} \textbf{2019}, \emph{1}, 3 032005.
	
	\bibitem{kariyado2017topological}
	T.~Kariyado, X.~Hu,
	\newblock \emph{Sci. Rep.} \textbf{2017}, \emph{7}, 1 16515.
	
	\bibitem{xu2016accidental}
	L.~Xu, H.-X. Wang, Y.-D. Xu, H.-Y. Chen, J.-H. Jiang,
	\newblock \emph{Opt. Express} \textbf{2016}, \emph{24}, 16 18059.
	
	\bibitem{jiang2019manipulation}
	Z.~Jiang, Y.-f. Gao, L.~He, J.-p. Sun, H.~Song, Q.~Wang,
	\newblock \emph{Phys. Chem. Chem. Phys.} \textbf{2019}, \emph{21}, 21 11367.
	
	\bibitem{zhou2021topological}
	R.~Zhou, H.~Lin, Y.~Liu, X.~Shi, R.~Tang, Y.~Wu, Z.~Yu, et~al.,
	\newblock \emph{Phys. Rev. A} \textbf{2021}, \emph{104}, 3 L031502.
	
	\bibitem{kim2020recent}
	M.~Kim, Z.~Jacob, J.~Rho,
	\newblock \emph{Light Sci. Appl.} \textbf{2020}, \emph{9}, 1 130.
	
	\bibitem{lan2023large}
	Z.~Lan, M.~L. Chen, J.~W. You, E.~Wei,
	\newblock \emph{Phys. Rev. A} \textbf{2023}, \emph{107}, 4 L041501.
	
	\bibitem{song2023observation}
	L.~Song, D.~Bongiovanni, Z.~Hu, Z.~Wang, S.~Xia, L.~Tang, D.~Song,
	R.~Morandotti, Z.~Chen,
	\newblock \emph{Adv. Opt. Mater.} \textbf{2023}, 2301614.
	
	\bibitem{poo2011experimental}
	Y.~Poo, R.-x. Wu, Z.~Lin, Y.~Yang, C.~T. Chan,
	\newblock \emph{Phys. Rev. Lett.} \textbf{2011}, \emph{106}, 9 093903.
	
	\bibitem{chen2020local}
	M.~L. Chen, L.~J. Jiang, Z.~Lan, E.~Wei,
	\newblock \emph{Opt. Express} \textbf{2020}, \emph{28}, 10 14428.
	
	\bibitem{wang20193d}
	B.~Wang, J.~Rodr{\'\i}guez, M.~A. Cappelli,
	\newblock \emph{Plasma Sources Sci. Technol.} \textbf{2019}, \emph{28}, 2
	02LT01.
	
	\bibitem{chen2017multiple}
	Z.-G. Chen, J.~Mei, X.-C. Sun, X.~Zhang, J.~Zhao, Y.~Wu,
	\newblock \emph{Phys. Rev. A} \textbf{2017}, \emph{95}, 4 043827.
	
	\bibitem{liu2022tunable}
	F.~Liu, Y.~Liu, Q.~Liu, Z.~Wu, Y.~Liu, K.~Gao, Y.~He, W.~Fan, L.~Dong,
	\newblock \emph{Plasma Sources Sci. Technol.} \textbf{2022}, \emph{31}, 2
	025015.
	
	\bibitem{wang2008reflection}
	Z.~Wang, Y.~Chong, J.~D. Joannopoulos, M.~Solja{\v{c}}i{\'c},
	\newblock \emph{Phys. Rev. lett.} \textbf{2008}, \emph{100}, 1 013905.
	
	\bibitem{liu2011molding}
	S.~Liu, W.~Lu, Z.~Lin, S.~Chui,
	\newblock \emph{Phys. Rev. B} \textbf{2011}, \emph{84}, 4 045425.
	
	\bibitem{gao2016photonic}
	W.~Gao, B.~Yang, M.~Lawrence, F.~Fang, B.~B{\'e}ri, S.~Zhang,
	\newblock \emph{Nat. Commun.} \textbf{2016}, \emph{7}, 1 12435.
	
	\bibitem{parker2020topological}
	J.~B. Parker, J.~Marston, S.~M. Tobias, Z.~Zhu,
	\newblock \emph{Phys. Rev. Lett.} \textbf{2020}, \emph{124}, 19 195001.
	
	\bibitem{fu2021topological}
	Y.~Fu, H.~Qin,
	\newblock \emph{Nat. Commun.} \textbf{2021}, \emph{12}, 1 3924.
	
	\bibitem{wu2018topologically}
	X.~Wu, F.~Ye, J.~M. Merlo, M.~J. Naughton, K.~Kempa,
	\newblock \emph{Adv. Opt. Mater.} \textbf{2018}, \emph{6}, 15 1800119.
	
	\bibitem{wu2023strong}
	J.~Wu, Y.~M. Qing,
	\newblock \emph{J. Chem. Phys.} \textbf{2023}, \emph{159}, 22.
	
	\bibitem{wang2020universal}
	C.~Wang, H.~Zhang, H.~Yuan, J.~Zhong, C.~Lu,
	\newblock \emph{Front. Optoelectron.} \textbf{2020}, \emph{13} 73.
	
	\bibitem{zhao2020first}
	R.~Zhao, G.-D. Xie, M.~L. Chen, Z.~Lan, Z.~Huang, E.~Wei,
	\newblock \emph{Opt. Express} \textbf{2020}, \emph{28}, 4 4638.
	
	\bibitem{yao2009edge}
	W.~Yao, S.~A. Yang, Q.~Niu,
	\newblock \emph{Phys. Rev. Lett.} \textbf{2009}, \emph{102}, 9 096801.
	
	\bibitem{ryu2010topological}
	S.~Ryu, A.~P. Schnyder, A.~Furusaki, A.~W. Ludwig,
	\newblock \emph{New J. Phys.} \textbf{2010}, \emph{12}, 6 065010.
	
	\bibitem{lan2021second}
	Z.~Lan, J.~W. You, Q.~Ren, E.~Wei, N.~C. Panoiu,
	\newblock \emph{Phys. Rev. A} \textbf{2021}, \emph{103}, 4 L041502.
	
\end{thebibliography}

\end{document}